# The role of dendritic spines in water exchange measurements with diffusion MRI: Double Diffusion Encoding and free-waveform MRI

Arthur Chakwizira[1], Kadir Şimşek[3,4], Filip Szczepankiewicz[1], Marco Palombo[3,4] and Markus Nilsson[2]

1. Medical Radiation Physics, Clinical Sciences Lund, Lund University, Lund, Sweden
2. Department of Clinical Sciences Lund, Radiology, Lund University, Lund, Sweden
3. Cardiff University Brain Research Imaging Centre (CUBRIC), School of Psychology, Cardiff University, Cardiff, United Kingdom,
4. School of Computer Science and Informatics, Cardiff University, Cardiff, United Kingdom

## Corresponding author:

Arthur Chakwizira
Department of Medical Radiation Physics, Lund University, Skåne University Hospital, SE-22185 Lund, Sweden
Email address: arthur.chakwizira@med.lu.se

**Word count:** 7500

## Sponsors/Grant numbers:

VR (Swedish Research Council)

- 2024-04968

eSSENCE

- 10:5

Hjärnfonden

- FO2024-0335-HK-73

Multipark
- This study was supported by MultiPark - A Strategic Research Area at Lund University

ALF
- The study was financed by Swedish governmental funding of clinical research (ALF).

Cancerfonden (The Swedish Cancer Society)

- 22 2011 Pj

UK Research and Innovation (UKRI)

- Future Leaders Fellowship MR/T020296/2

## Abstract

Time-dependent diffusion MRI enables the estimation of water exchange rates *in vivo*, yet reported values in grey matter remain inconsistent. While most studies attribute these estimates to membrane permeability, non-permeative geometric exchange has also been proposed. The present study investigates the contribution of geometric exchange between dendritic spines and shafts to diffusion MRI-derived exchange estimates. Monte Carlo simulations were performed in synthetic dendrites with varying spine morphology, density, and membrane permeability. Diffusion-weighted signals were generated using multiple protocols—including single diffusion encoding, double diffusion encoding, and free waveforms—and were analysed using four frameworks: the Kärger model (via kurtosis time-dependence), correlation tensor imaging, Restriction-Exchange, and Multi-Gaussian Exchange with transient kurtosis (tMGE). Dendritic spines were found to impart similar time-dependence signatures on the diffusion-weighted signal as permeative exchange (signal decrease with diffusion time). The effect was modulated by both spine morphology and density. Both the exchange rate and microscopic kurtosis increased with spine density. The tMGE method demonstrated the ability to disentangle geometric from permeative exchange. Non-permeative exchange in dendritic spines has a non-negligible impact on exchange estimates obtained with diffusion MRI and should be considered in future studies. Diffusion MRI exchange estimates may provide a non-invasive proxy for dendritic spine density, with potential applications in studies of neurological disorders.

# 1 Introduction

Time-dependent diffusion MRI exhibits unparalleled sensitivity to the microstructure of biological tissue and has been applied to measure water exchange rates in the brain [1–4]. Many studies have used filter-exchange imaging (FEXI) [5–7] to measure exchange in the healthy human brain [8], in brain tumours [2], and in animal models [9]. Other studies have utilised diffusion-exchange spectroscopy (DEXSY) to measure water exchange rates in the mouse spinal cord [10,11]. These approaches rely on double-diffusion encoding (DDE [12]), while other methods combine the standard model [13] and the Kärger model [14] and leverage single diffusion encoding (SDE[12]) acquisitions with variable diffusion times [4,15,16]. Yet other studies exploit cumulant-expansion analyses tracking the diffusion time dependence of the diffusional kurtosis [9,17,18]. Recent work proposed a unification of all approaches for measuring exchange with diffusion MRI by defining a generalised exchange sensitivity for arbitrary gradient waveforms [19,20]. This approach was recently used to map exchange in the healthy brain, unconfounded by effects of restricted diffusion [3]. As evidenced by these innovations, exchange measurement with diffusion MRI is an area of highly active research.

Despite recent advancements, reliably estimating exchange with diffusion MRI remains a challenge—particularly in grey matter. In healthy white matter, there is consensus in the field that exchange is slow and negligible at clinically accessible diffusion times, primarily due to the low permeability of myelinated axons [21–25]. In contrast, ischemic white matter has been shown to exhibit non-negligible exchange [1,26]. In grey matter, evidence suggests that exchange between neurites and the extracellular space is non-negligible [16,25,27]. However, reported exchange estimates in grey matter are highly variable, ranging from a few to hundreds of milliseconds [16].

Discrepancies in exchange rates measured with diffusion MRI can be attributed to several factors. The first regards experimental design, where different methods use different acquisition approaches and likely probe different exchange regimes. Single diffusion encoding with variable diffusion times may conflate exchange with effects of restricted diffusion, as shown in previous work [20]. Some studies are also performed on fixed tissue which changes the membrane permeability [16,28]. The second factor concerns modelling challenges. While the Kärger model is the mainstay of all exchange estimation with diffusion MRI, it makes strong assumptions about Gaussian diffusion in each compartment and barrier-limited exchange, which may be violated in brain tissue. Mechanisms of exchange in grey matter are generally underexplored [29]. Recent work has demonstrated that exchange contrasts in diffusion MRI can emerge in the absence of membrane permeation, as long as water diffuses between distinct environments [30].

In this work, following previous studies [31–35], we investigate yet another potential mechanism of exchange in grey matter: dendritic spines. These are small protrusions on the surface of dendrites that form functional contacts with axons of neighbouring neurons. Spines were first observed by Ramón y Cajal in 1888 using Golgi staining and confirmed fifty years later by electron microscopy as the primary postsynaptic site for excitatory synapses [36,37]. Ubiquitously expressed throughout the mammalian brain, dendritic spines manifest in multiple shapes reflecting their level of maturity, with filopodia-like spines being the least stable and mushroom-shaped ones being the most stable [38,39]. We hypothesise that the diffusive water transport between dendritic shafts and spines imparts exchange-like signatures on the diffusion-weighted signal and may, alongside permeative exchange across highly permeable cell membranes, account for the high exchange rates reported in grey matter. We test this hypothesis by performing extensive Monte Carlo simulations in substrates mimicking dendritic spines with variable morphology and density. Our results indicate that intracellular diffusive exchange in dendritic spines is a plausible explanation for the high exchange rates estimated by diffusion MRI in grey matter.

## 2 Theory

### 2.1 Narrow escape theory

The diffusional coupling between the dendritic spine and shaft has been investigated extensively in previous studies seeking to understand synaptic plasticity and receptor dynamics [40–48]. Such studies have focused on theory development and Monte Carlo simulations, or employed two-photon photobleaching experiments to track the equilibration of molecules in dendritic spines. Expressions describing the one-directional diffusive exchange rates between the spine and shaft have been proposed, with varying assumptions. This theory, summarised below, is now generally known as the "narrow-escape problem", alluding to the diffusion of a particle fully confined to a compartment except for a narrow and localised exit.

*Spine-to-shaft exchange rate*

Given a dendritic spine comprising a spherical head and cylindrical neck (Fig. 1A), the mean residence time for particles initially in the spine and diffusing out to the shaft without return is given by [41,49,50]

$$\tau_{spine} = \frac{1}{p_0}\tau_{head} + \tau_{neck} + \frac{1 - p_0}{p_0}\tau_{return} \, , \qquad (1)$$

where $p_0$ is the probability that a particle having entered the neck does not return to the head, $\tau_{head}$ and $\tau_{neck}$ are the mean residence times in the spine head and neck, respectively, and $\tau_{return}$ is the average time a particle spends in the head upon its first return from the neck before eventually exiting to the shaft. The probability and residence times in Eq. 1 depend on the spine morphology according to

$$p_0 = \frac{\alpha r_{neck}}{l_{neck}} \, , \qquad (2)$$

$$\tau_{head} = \frac{V}{4 r_{neck} D_0}\left[1 + \frac{r_{neck}}{r_{head}}\ln\left(\frac{r_{head}}{r_{neck}}\right)\right] \, , \qquad (3)$$

$$\tau_{neck} = \frac{l_{neck}^2}{6 D_0} \, , \qquad (4)$$

and

$$\tau_{return} = \frac{\alpha r_{neck} l_{neck}}{3 D_0} \, , \qquad (5)$$

where $V$ is the spine head volume, $D_0$ is the bulk diffusivity, $r_{neck}$ is the spine neck radius (see Fig. 1A), $r_{head}$ is the spine head radius, $l_{neck}$ is the spine neck length and $\alpha$ is a constant such that a particle is considered in the neck when it has travelled a distance of $\alpha r_{neck}$ into it. It can be shown that when the spine neck is small compared to the radius of the head, the diffusion process above is captured by a single exponential of rate $k_{spine \rightarrow shaft} = 1/\tau_{spine}$. Using this, setting $\alpha = 1$ and inserting $V = \frac{4}{3}\pi r_{head}^3$ yields

$$k_{spine \to shaf} = \frac{6D_0 r_{neck}^2}{l_{neck} \cdot \left( 2\pi r_{head}^3 + 3r_{neck}^2 l_{neck} + 2\pi r_{head}^2 r_{neck} \ln\left(\frac{r_{head}}{r_{neck}}\right) - 2r_{neck}^3 \right)}. \quad (6)$$

Equation 6 suggests that the spine-to-shaft exchange rate decreases with the neck length and head radius and increases with the neck radius.

*Shaft-to-spine exchange rate*

The narrow-escape problem for diffusion from the shaft to the spine has recently been studied by representing the shaft as a straight cylinder featuring many circular pores (openings) on its surface [48]. For small pore densities, the exchange rate from the shaft to the spines is well-approximated by the inverse of the mean residence time in the shaft, and is given by

$$k_{shaft \to spine} = k_{max} \cdot \frac{p}{p + 2\sigma(1-p)(2-\sigma)} , \quad (7)$$

where $k_{max}$ is the maximal exchange rate for a fully absorbing cylinder, $p$ and $\sigma$ are dimensionless parameters. These quantities relate to the pore density and geometry via

$$k_{max} = \frac{8D_0}{r_{shaft}^2} , \quad (8)$$

$$p = \frac{N r_{neck}^2}{2 r_{shaft} L_{shaft}} , \quad (9)$$

and

$$\sigma = \frac{\pi}{4} \frac{r_{neck}}{r_{shaft}} , \quad (10)$$

where $r_{shaft}$ and $L_{shaft}$ are the shaft radius and length and $N$ is the number of spines over the length $L_{shaft}$. Defining the spine density as $\rho_{spine} = N/L_{shaft}$, Eq. 7 simplifies to

$$k_{shaft \to spine} = \frac{64\rho D_0 r_{neck}}{2\pi r_{shaft}^2 (8 - \pi r_{neck}) + \rho \left( 8 r_{neck} r_{shaft}^2 - 8\pi r_{shaft} r_{neck}^2 + \pi^2 r_{neck}^3 \right)}. \quad (11)$$

Equation 11 is valid under narrow-escape conditions, where the neck radius is much smaller than the shaft radius ($r_{neck} \ll r_{shaft}$).

## 2.2 Signal representations

The impact of dendritic spines on diffusion MRI measurements is appropriately studied using a variety of different methods. This section provides a brief overview of the approaches used to estimate diffusional kurtosis and exchange rates—parameters expected to be influenced by spines.

*Time-dependent kurtosis*

Diffusional kurtosis in realistic structures featuring both restricted diffusion and exchange has a non-monotonic dependence on the diffusion time, first increasing due to restriction and later decreasing due to exchange [9,17,51,52]. The kurtosis is estimated at each diffusion time via the signal representation

$$\ln(S/S_0) \approx -b\overline{D} + \frac{1}{6}b^2\overline{D}^2 K \,, \tag{12}$$

where $S_0$ is the signal in the absence of diffusion weighting, $b$ is the b-value, $\overline{D}$ is the mean diffusivity and $K$ is the kurtosis. The exchange-driven decay of the kurtosis is described by the Kärger model via [14,19,51,53]

$$K(t_d) = K_0 \cdot \frac{2}{kt_d}\left[1 - \frac{1}{kt_d}(1 - e^{-kt_d})\right] + K_\infty, \tag{13}$$

where $t_d$ is the diffusion time, $K_0$ is the initial kurtosis, $k$ is the exchange rate and $K_\infty$ captures residual kurtosis contributions due to, for example, the presence of non-exchanging compartments.

*Restriction-Exchange (ResEx)*

In order to disentangle the influence of restricted diffusion and exchange, the so-called ResEx (Restriction-Exchange) approach can be used. It leverages free gradient waveforms (FWF) to gain selective sensitivity to restricted diffusion and exchange [3,19,20,54]. Assuming an acquisition protocol designed to null restriction-driven time-dependence, the signal is given by

$$\ln(S/S_0) \approx -b\overline{D} + \frac{1}{6}b^2\overline{D}^2 K_0 h(k) \,, \tag{14}$$

where $h(k)$ is the exchange-weighting function given by

$$h(k) = \frac{2}{b^2} \int_0^T e^{-kt} \int_0^T q^2(t) q^2(t+t') \, dt' dt \approx 1 - k\Gamma, \qquad (15)$$

where $q(t) = \gamma \int_0^t g(\tau) d\tau$ is the dephasing q-vector, $g$ is the gradient waveform of duration $T$ and $\Gamma = \frac{2}{b^2} \int_0^T t \int_0^T q^2(t) q^2(t+t') \, dt' dt$ is the exchange-weighting time (or exchange sensitivity). The approximation in Eq. 15 holds for small values of $kt$.

*Correlation Tensor Imaging (CTI)*

Exchange may also be treated as a source of microscopic kurtosis and studied using correlation tensor imaging (CTI) [55,56] which leverages SDE and DDE in the long mixing time regime to resolve different kurtosis sources, according to

$$\ln(S/S_0) \approx -b\overline{D} + \frac{1}{6} b^2 \, \overline{D}^2 \big( K_I + b_\Delta^2 K_A + b_\mu^2 K_\mu \big), \qquad (16)$$

where $K_I$, $K_A$ and $K_\mu$ are the isotropic, anisotropic and microscopic kurtoses, respectively, $b_\Delta^2 = \frac{b_1^2 + b_2^2 + b_1 b_2 (3\cos^2\theta - 1)}{(b_1 + b_2)^2}$ where $b_1$ and $b_2$ are the b-values of the first and second diffusion encoding blocks and $\theta$ is the angle between them, and, finally, $b_\mu^2 = \frac{b_1^2 + b_2^2}{(b_1 + b_2)^2}$. Thus, anisotropic kurtosis is obtained by contrasting parallel and orthogonal DDE measurements, and microscopic kurtosis by contrasting SDE and parallel DDE signals at the same b-value.

*Multi-Gaussian exchange with transient kurtosis (tMGE)*

A recently proposed analysis framework—multi-Gaussian exchange with transient kurtosis (tMGE)—combines insights from both ResEx and CTI and leverages SDE and DDE with variable mixing times to decompose the microscopic kurtosis into its two constituents: exchange and so called "transient kurtosis" [57]. The signal equation is

$$\ln(S/S_0) \approx -b\overline{D} + \frac{1}{6} b^2 \overline{D}^2 h(k, \Delta, t_m) \big[ K_I^0 + h_\Delta^2(k, \Delta, t_m) K_A^0 \big] + \frac{1}{6} b^2 \overline{D}^2 \big[ K_I^\infty + b_\Delta^2 K_A^\infty + b_\epsilon^2 K_\epsilon \big], (17)$$

where $K_I$, $K_A$ and $K_\epsilon$ are the isotropic, anisotropic and transient kurtoses, respectively, superscripts "0" and "∞" denote initial and long-time, respectively, $b_\epsilon = b_\mu$, $h$ and $h_\Delta$ are the isotropic and anisotropic projections of the exchange-weighting tensor given by

$$\mathbb{H}(k) = 2 \int_0^T e^{-k} \int_0^T \mathbf{q}^{\otimes 2}(t') \otimes \mathbf{q}^{\otimes 2}(t'+t) \, \mathrm{d}t' \, \mathrm{d}t. \tag{18}$$

# 3 Methods

## 3.1 Simulation substrate design

A dendritic spine model featuring a cylindrical shaft, cylindrical spine necks and spherical spine heads was used in this study (Fig. 1A). This design mimicked dendrites with stable mushroom-shaped spines, the most common type in mature neurons [58]. The shaft diameter was fixed at 1 μm and its total length set to 250 μm within the repeating unit of the substrate. Spine density was varied by randomly scattering spines along the surface of the cylindrical shaft, ensuring no overlaps between spine heads or necks. While maintaining a fixed density and location of the spines, the spine neck length, neck diameter and head diameter were varied to investigate variations in morphology. These morphology variations were based on literature values shown in Table 1 [59] and resulted in a total of 28 unique realisations of dendrites with spines.

Table 1: Dendritic spine morphology parameters used in simulations.

| Neck length [μm] | Neck diameter [μm] | Head diameter [μm] | Spine density [μm⁻¹] |
|---|---|---|---|
| 0.75, 1.0, 1.25, 1.5, 1.75, 2.0 | 0.2 | 0.9 | 2 |
| 1.5 | 0.05, 0.1, 0.2, 0.3, 0.4, 0.5 | 0.9 | 2 |
| 1.5 | 0.2 | 0.6, 0.7, 0.8, 0.9, 1.0, 1.2 | 2 |
| 1.5 | 0.2 | 0.9 | 0, 0.2, 0.4, 0.6, …, 2 |

The substrate design was done using in-house written MATLAB code available at github.com/arthur-chakwizira/spines. To investigate the interplay between membrane permeability and dendritic spines, substrates with variable spine density (bottom row in Table 1) were mixed with extracellular water and permeability was adjusted to yield exchange rates of 0, 25 and 50 s⁻¹. The method used to determine the relationship between permeability and exchange rate is described below.

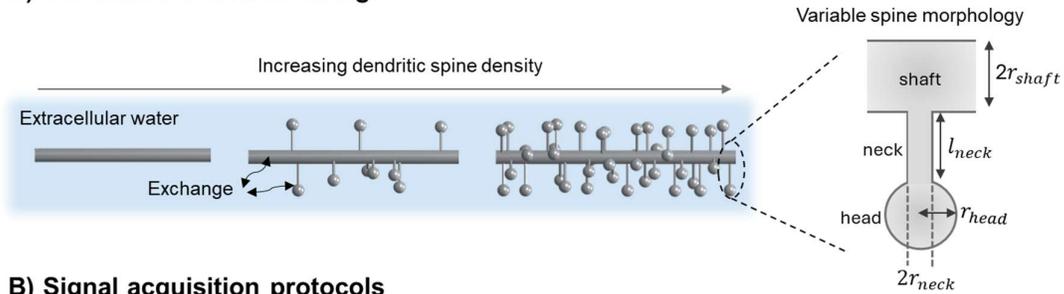

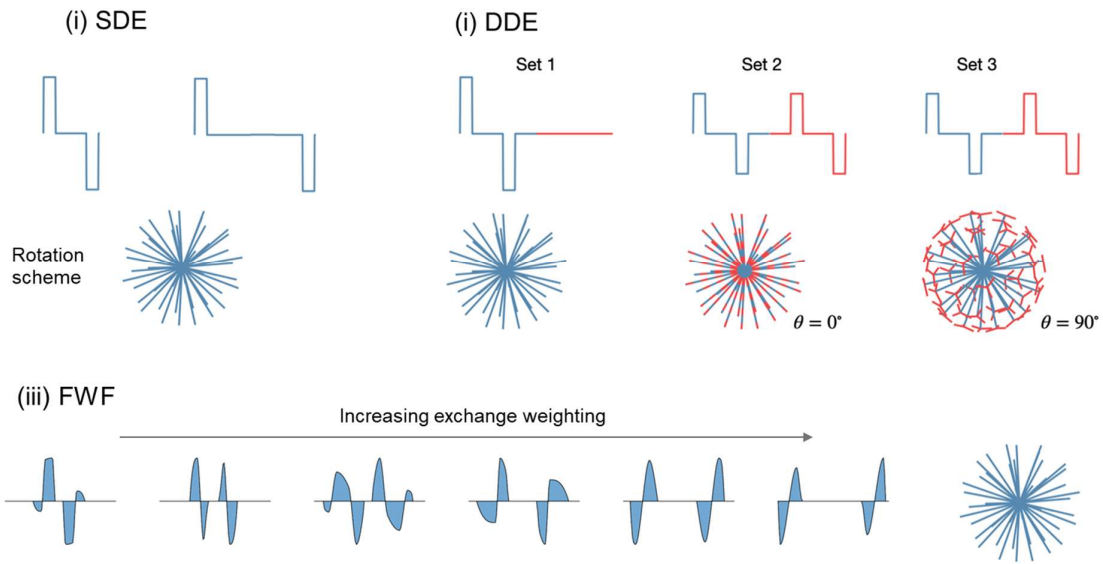

Figure 1: Simulation substrates and acquisition protocols for studying the effect of dendritic spines on diffusion MRI exchange estimates. Panel (A) shows a dendrite model comprising a cylindrical shaft, cylindrical spine necks and spherical spine heads. Spine density was varied by progressively placing more spines along the shaft. Spine morphology was also varied by changing the neck length and diameter and the head diameter. Dendrites were also mixed with 20% extracellular water and allowed to exchange with it at varying rates. Panel B shows the three types of protocols used in this study: (i) SDE with variable diffusion times rotated in 20 directions chosen to minimise electrostatic repulsion on a sphere (ii) DDE featuring SDE, parallel and orthogonal DDE with variable mixing times, rotated as shown and (iii) FWF with varying exchange weighting. Parameters for the different protocols are shown in Table 2.

## 3.2 Protocol design

Diffusion-weighted signals were simulated using three different types of protocols: SDE with variable diffusion times and pulse widths, DDE with variable mixing times and free waveforms (FWF) with variable exchange weighting (Fig. 1B). The variety of protocols enabled the study of the effects of dendritic spines on the parameters obtained with a wide range of methods outlined in Section 2.2. All simulated signals and protocols will be made available online at github.com/arthur-chakwizira/spines.

*Single Diffusion Encoding (SDE)*

The SDE type featured four variants. First, Protocol I with $\delta = 0.5$ ms and fifteen diffusion times between 0.5 and 99.3 ms for studying agreement between simulated and theoretical exchange rates.

Second, Protocol II with $\delta = 1$ ms and fifteen diffusion times between $t_{d,min}$ and 99.3 ms where $t_{d,min} = 3, 10$ and 15 ms for studying the effect of varying the range of diffusion times. Third, Protocol III with $\delta = 1, 3$ and 5 ms with fifteen diffusion times evenly distributed between 3.5 ms and the maximum possible value ($T - 4\delta/3$ where $T$ = 100 ms) for studying the effect of varying the pulse width. Fourth, and finally, Protocol IV with $\delta = 9$ ms and fifteen diffusion times in the range 6.5 to 88 ms for illustrating effects of hardware constraints. The first three protocols were designed for arbitrary hardware, while the fourth obeyed a maximum gradient strength of 300 mT/m and slew rate of 200 T/m/s, typical of current state-of-the-art *in vivo* scanners, such as the head-only MAGNUS system [60]. Diffusion times followed an exponential sampling pattern for all four protocols to capture kurtosis time-dependence at short times due to the high exchange rates expected in dendritic spines. All four protocols used one-dimensional gradient waveforms with b-values [0, 0.5, 1, 1.5, 2.5] ms/μm², rotated in 20 directions per b-value chosen to minimise electrostatic repulsion on a sphere (Fig. 1 B (i)).

*Double Diffusion Encoding (DDE)*

Three DDE-type protocols were used in this study. The first (Protocol V), was designed under the constraints of the MAGNUS system above and featured $\delta = 8$ ms, $\Delta = 12$ ms and $t_m = 36$ ms. Three sets of waveforms were generated following previous work [56,57].

Set 1 comprised an SDE using the same b-values and rotation scheme as the SDE protocols above. Set 2 comprised parallel DDE with the same total b-values as the SDE, divided equally between the first and second encoding blocks, and rotated as in Set 1. Set 3 featured orthogonal DDE with the same b-values as in Set 2. The first encoding block was rotated as in Set 2 and for each direction, the second encoding block was rotated in three perpendicular directions, yielding in total 60 rotations per b-value for Set 3. The purpose of this protocol was to enable characterisation of dendritic spines using CTI, as described in Section 2.2.

To investigate whether effects of spines could be separated from effects of permeative exchange using tMGE, an additional DDE protocol (Protocol VI) was designed using $\delta = 0.5$ ms, $\Delta = 1$ ms and multiple mixing times for both the parallel and orthogonal DDE: $t_m = [0.5, 1, 2, 4, 8, 36, 64, 100, 150, 200]$ ms. The high sampling density at short mixing times was meant to capture short-time kurtosis dynamics induced by fast exchange. This protocol used the same b-values and rotation schemes as Protocol V, and no hardware constraints. No constraints were placed on this protocol because its purpose was to investigate whether tMGE in principle can separate geometric from permeative exchange. Aiming to investigate the feasibility of tMGE on realistic hardware, another DDE protocol (Protocol VII) was designed to be executable on the MAGNUS system. It had $\delta = 8$ ms, $\Delta = 12$ ms and the same mixing times, b-values and rotation schemes as Protocol VI.

*Free waveforms (FWF)*

The FWF protocol was adapted from previous work [3] and featured six waveforms with increasing sensitivity to exchange ($\Gamma$) and fixed sensitivity to restricted diffusion (Fig. 1B (iii)). The protocol was designed for the MAGNUS system, featured a maximum waveform duration of 100 ms and used b-values of [0 1 2 3 4] ms/$\mu$m$^2$ with 20 rotations per b-value as for the SDE above. All protocols described in this section are summarised in Table 2.

Table 2: Protocols used in simulations

| Protocol name | | Timing parameters | B-values [ms/µm²] | Hardware constraints |
|---|---|---|---|---|
| SDE | Protocol I | $\delta = 0.5$ ms<br>$t_d = 0.5, \dots, 99.3$ ms | 0, 0.5, 1, 1.5, 2.5 | None |
| | Protocol II | $\delta = 1$ ms<br>$t_d = t_{d,min}, \dots, 99.3$ ms<br>$t_{d,min} = [3, 10, 15]$ ms | 0, 0.5, 1, 1.5, 2.5 | None |
| | Protocol III | $\delta = [1, 3, 5]$ ms<br>$t_d = 3.5, \dots, T - 4\delta/3$ | 0, 0.5, 1, 1.5, 2.5 | None |
| | Protocol IV | $\delta = 9$ ms<br>$t_d = 6.5, \dots, 88$ ms | 0, 0.5, 1, 1.5, 2.5 | 300 mT/m, 200 T/m/s |
| DDE | Protocol V | $\delta = 8$ ms, $\Delta = 12$ ms, $t_m = 36$ ms | 0, 0.5, 1, 1.5, 2.5 | 300 mT/m, 200 T/m/s |
| | Protocol VI | $\delta = 0.5$ ms, $\Delta = 1$ ms, $t_m = [$ 10, 15, 30, 50, 70, 100, 150, 200] ms | 0, 0.5, 1, 1.5, 2.5 | None |
| | Protocol VII | $\delta = 8$ ms, $\Delta = 12$ ms, $t_m = [$ 10, 15, 30, 50, 70, 100, 150, 200] ms | 0, 0.5, 1, 1.5, 2.5 | 300 mT/m, 200 T/m/s |
| FWF | Protocol VIII | $\Gamma = [5, 12, 19, 26, 33, 39]$ ms | 0,1,2,3,4 | 300 mT/m, 200 T/m/s |

## 3.3 Numerical simulations

Monte Carlo simulations were performed using an inhouse-written GPU-accelerated framework described previously [7,20,57] (github.com/arthur-chakwizira/Pasidi). All simulations used periodic boundary conditions, a bulk diffusivity of 2 µm²/ms and temporal resolution of 0.1 µs.

*Measurement of spine-to-shaft and shaft-to-spine exchange rates from trajectories*

The first set of simulations aimed to measure the one-directional spine-to-shaft and shaft-to-spine exchange rates for comparison with the predictions of Eq. 6 and 11. Ten thousand particles were initialised uniformly in the spines and allowed to diffuse out to the shaft, with each particle that had left its spine being eliminated from the simulation. To limit the effect of non-mono-exponential exchange, the simulation was terminated when half of all particles had left (c.f supplementary Fig. A1). The spine-to-shaft exchange rate was determined by fitting the following equation to the spine population ($N$)

$$N(t) = N(0)e^{-k_{spine \to shaft} \cdot t}. \tag{19}$$

The simulation was repeated with the ten thousand particles being initialised in the shaft instead and allowed to diffuse out to the spines, allowing estimation of $k_{shaft \to spine}$. This procedure was repeated for all twenty-eight dendrite realisations in Table 1 to investigate the dependence of the exchange rates above on spine morphology and density.

*Intracellular signal generation*

The next set of simulations used 1 million particles constrained to the intracellular environment (shaft, neck, and spine). In each of the twenty-eight dendrite realisations in Table 1, diffusion-weighted signals were generated using all protocols described in Section 3.2 (and shown in Table 2).

*Incorporating exchange with the extracellular space*

The geometry of spines makes close-packing of dendrites difficult, which entails that the repeating unit of the substrate is predominantly extracellular space. A very low intracellular volume fraction makes exchange simulations impractical, because the particle density must be kept equal in both intra- and extracellular spaces. To resolve this problem, we designed a simulation environment where intracellular particles encountering the membrane could move to the extracellular space with a probability governed by the set permeability. Once in the extracellular space, however, the diffusion process became completely Gaussian, meaning the particle no longer "saw" the dendrite. At each step in the extracellular space, the particle could return to the intracellular space with a probability governed by the desired exchange rate. If the particle was allowed to return, a new dendrite appeared in its position, and the particle found itself in a random position inside the dendrite. This approach enabled a high virtual packing of dendrites, at the expense of reduced plausibility of the diffusion process in the extracellular space.

The relationship between the permeability of the dendrite and the exchange rate to the extracellular space, $k_{in \to ex}$, was determined by running a leakage simulation where all particles were initialised in the dendrite and allowed to diffuse to the extracellular space without return. The exchange rate was obtained by fitting a mono-exponential decay (Eq. 19) to the resulting dendrite population-versus-time curve, following the approach of [61]. This was repeated for a range of different permeabilities, resulting in a calibration curve connecting permeability to exchange rate. It was necessary to determine the relationship between exchange rate and permeability in this way instead of via the surface-to-volume ratio [62] because the latter tends to underestimate the true permeability in irregular geometries. Calibration curves were obtained for ten different dendrites with varying spine densities (bottom row of Table 1). Focus was limited to these substrates since the goal was to investigate the interplay between dendritic spines and permeative intra-extracellular exchange.

The probability of exiting the dendrite was related to the permeability ($\kappa$) through [63,64]

$$p_{in \to ex} = 4\kappa \sqrt{\frac{\Delta t}{2nD_0}} \ , \qquad (20)$$

where $\Delta t$ is the simulation time step, $D_0$ is the bulk diffusivity and $n = 3$ is the number of spatial dimensions. For a desired total exchange rate $k = k_{in \to ex} + k_{ex \to in}$, while $k_{in \to ex}$ was determined by the permeability above, $k_{ex \to in}$ was set to

$$k_{ex \to in} = \frac{k}{1 + f_{ex}/f_{in}} \ , \qquad (21)$$

where $f_{in}$ and $f_{ex} = 1 - f_{in}$ are the dendrite and extracellular signal fractions, respectively. In this work, we used $f_{in} = 0.8$. Equation 21 ensured that the mass balance condition $f_{ex}k_{ex \to in} = f_{in}k_{in \to ex}$ was satisfied [14,19,53]. The probability of transition from the extracellular space to the dendrite was then given by

$$p_{ex \to in} = f_{in}(1 - e^{-k\Delta t}) \,. \tag{22}$$

The dendrite permeability was varied to yield three exchange rates, $k = 0, 25$ and $50$ s$^{-1}$. Signals were generated at these three exchange rates in all ten substrates of variable spine density, using the DDE and FWF protocols.

### 3.3 Data analysis

All signals simulated in this study were powder-averaged prior to fitting. To study the signature of dendritic spines on the time-dependence of the diffusion-weighted signal and to compare the intracellular exchange rates predicted by theory to those estimated under the same conditions by diffusion MRI, signals generated with Protocol I were used. Equation 12 was fitted to the powder-averaged signal at each diffusion time, yielding $\bar{D}(t_d)$ and $K(t_d)$ curves. Eq. 13 was then fitted to the decreasing part of the $K(t_d)$ curve, providing the exchange rate $k$ incorporating contributions from both $k_{shaft \to spine}$ and $k_{spine \to shaft}$.

A similar analysis was performed for signals acquired with Protocol II and III to study the effect of varying ranges of diffusion time and pulse width on the exchange signatures from dendritic spines. The analysis was limited to substrates with fixed morphology and varying spine density.

Signals generated with protocols designed for a 300 mT/m system were used to assess the exchange imprint of dendritic spines under more realistic measurement settings. Signals obtained with Protocol IV were analysed as described above, yielding $\bar{D}(t_d)$, $K(t_d)$ and the exchange rate $k$. FWF signals (Protocol VIII) were analysed by initially fitting Eq. 12 to signals acquired with each gradient waveform, providing $\bar{D}(\Gamma)$ and $K(\Gamma)$ curves. Eq. 14 was then fitted to the signals to obtain the exchange rate, $k$, as a function of spine density. For DDE (Protocol V), a fit of Eq. 12 was also initially performed to give $\bar{D}(.)$ and $K(.)$, at a fixed mixing time of 36 ms for both parallel and orthogonal DDE. The CTI signal representation (Eq. 16) was then fitted to the signals to obtain microscopic kurtosis estimates as a function of spine density.

To explore how permeative exchange with the extracellular space influences the intracellular exchange signatures from dendritic spines, the CTI and FWF analyses described above were also applied to powder-averaged signals obtained in the presence of intra-extracellular exchange at rates of 0, 25 and 50 $s^{-1}$. Finally, an investigation applying tMGE to powder-averaged DDE signals acquired with multiple mixing times (Protocol VI and VII) was also performed. This yielded estimates of transient kurtosis and exchange rate as a function of spine density. To study robustness of tMGE to noise, the analysis using Protocol VI was repeated after adding Rice-distributed noise to the signals at a generous signal-to-noise ratio (SNR) of 200.

# 4 Results

Simulation results showing the relationship between dendritic spine morphology and water exchange rates are presented in Fig. 2. The spine-to-shaft exchange rates decrease with the neck length, increase with neck diameter, decrease with head diameter and are insensitive to spine density. The measurements are in good agreement with theory (dashed lines). Shaft-to-spine exchange rates are an order of magnitude higher, and are independent of neck length and head diameter but increase with neck diameter and spine density. The trends align with the theoretical predictions. There is a small systematic underestimation of the true exchange rate in most of the simulations.

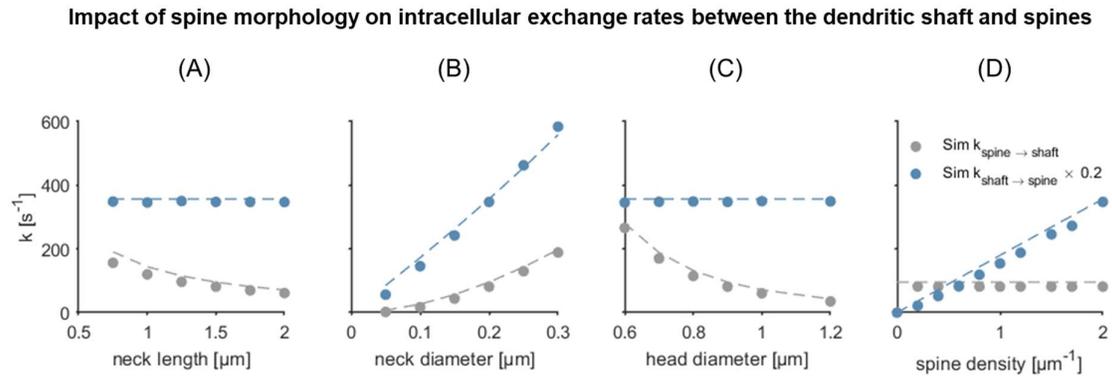

Figure 2: Exchange rates as a function of spine density and morphology. The main result is that spine density and morphology are important modulators of intracellular exchange inside spiny dendrites. Dots show estimates from simulations (Eq. 19) and dashed lines represent theoretical predictions (Eq. 6 and 11). Spine-to-shaft exchange rates decrease with neck length (A) and head diameter (C), increase with neck diameter (B) and are independent of spine density (D). Shaft-to-spine exchange rates are independent of neck length (A) and head diameter (C) but increase with neck diameter (B) and spine density (D). Note that the shaft-to-spine exchange rates (both simulated and predicted) are scaled by 0.2 to bring them to the same scale as the spine-to-shaft rates, for better visualisation.

Figure 3 illustrates the impact of dendritic spine density and morphology on the mean diffusivity (MD), kurtosis ($K_T$) and the exchange rate estimated from the time dependence of the diffusional kurtosis. The mean diffusivity generally decreases with diffusion time, which is a characteristic of restricted diffusion. Furthermore, the kurtosis generally decreases with diffusion time—a signature of diffusional exchange. Signal-vs-b curves also highlight an exchange signature (signal decrease with diffusion time) and can be found in Fig. A2 of the supplementary. At very high exchange rates (such as the case with the largest neck diameter and smallest head diameter), the kurtosis initially undergoes rapid decay and then increases towards an equilibrium where it remains constant with diffusion time. Estimated exchange rates show a strong dependence on both spine morphology and density: decreasing with the neck length and head diameter and increasing with the neck diameter and spine density. Since the estimated exchange rate contains contributions from both the forward and reverse rates, these trends are in good agreement with theory (Fig. 2). However, the absolute values are notably lower than theoretical expectations.

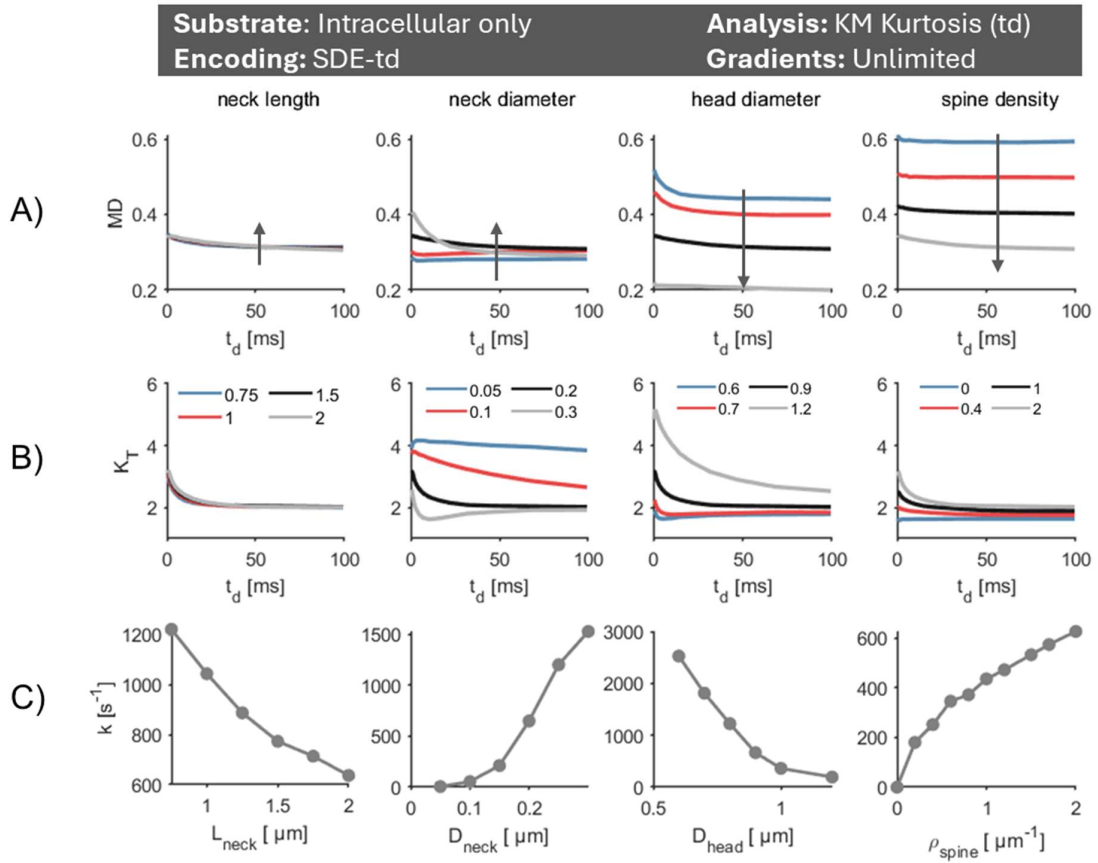

Figure 3: Influence of spine morphology on dMRI parameter estimates from simulations using Protocol I with intracellular particles only. Panel (A) shows that the mean diffusivity (MD) is largely independent of neck length, increases with neck diameter and head diameter, and decreases with spine density. In all cases, MD decreases slightly with diffusion time. Panel (B) shows that the mean kurtosis ($K_T$) is largely independent of neck length, decreases with neck and head diameter, and increases with spine density. The kurtosis generally decreases with diffusion time, normally interpreted as an indication of exchange. Panel (C) shows that the total exchange rate decreases with neck length and head diameter and increases with neck diameter and spine density, all in agreement with theoretical expectations. Taken together, these results show that intracellular diffusion in dendritic spines carries signatures of diffusional exchange.

The impact of varying the minimum diffusion time and the gradient pulse width on the exchange rates estimated from the kurtosis time-dependence is shown in Fig. 4. The minimum accessible diffusion time strongly determines the sensitivity of the exchange rate to changes in spine density, with shorter diffusion times giving sensitivity over a broader range of densities. Increasing the pulse width does not alter the sensitivity of the exchange rate to spine density but causes a systematic decline in the estimated values. $K(t_d)$ curves used to obtain the estimates in Fig. 4 can be found in supplementary Fig. A3. Overall, Fig. 4 shows that the relationship between exchange rates estimated by dMRI and the dendritic spine density is heavily influenced by the protocol (and thus the available hardware).

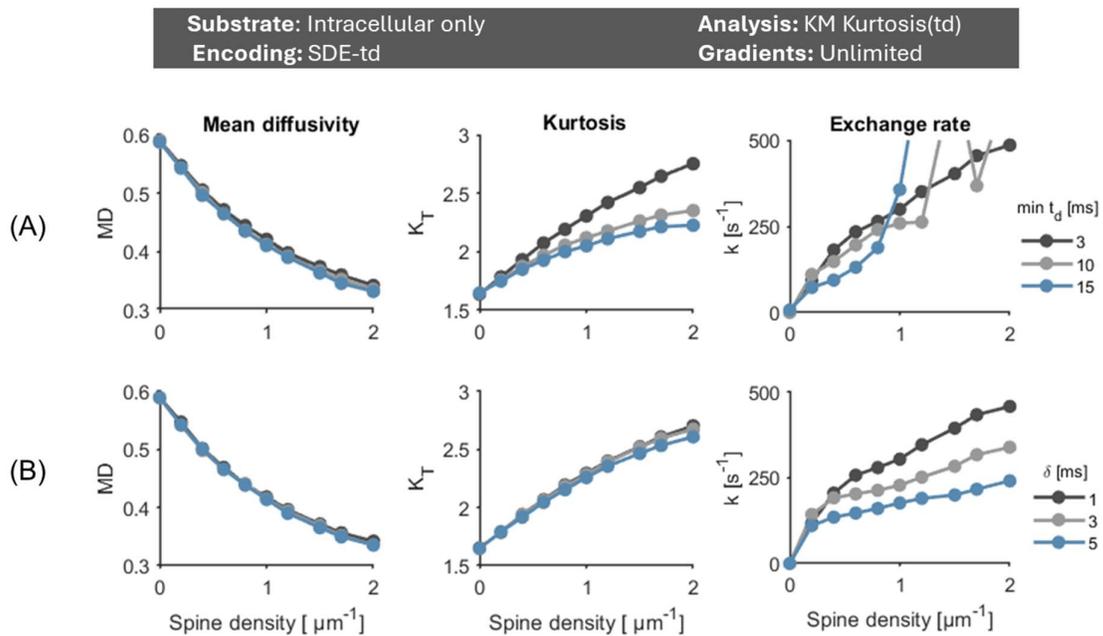

Figure 4: Protocol-dependence of intracellular exchange rates in dendritic spines. Results are shown for Protocol II and III applied to intracellular particles, with exchange estimation done using Eq. 13. Panel A shows the effect of increasing the minimum available diffusion time. Protocols featuring short diffusion times exhibit sensitivity to a broader range of spine densities. With longer diffusion times, sensitivity to high spine densities (which give fast exchange) is lost. Panel B shows the effect of varying pulse width. At low spine densities, the pulse width has no effect but causes a bias in exchange estimates at higher densities. The main message of this figure is that the effect of dendritic spines on diffusion MRI-estimated exchange rates is protocol-dependent, and that accurate characterisation requires high performance gradients that enable short diffusion times and narrow pulses.

Figure 5 shows the mean diffusivity, kurtosis and exchange estimates obtained with SDE $K(t_d)$, ResEx (FWF) and CTI, as a function of dendritic spine density. The SDE mean diffusivity shows a subtle time-dependence, while the kurtosis time-dependence is more pronounced, suggesting that exchange effects from dendritic spines persist for realistic SDE acquisitions (300 mT/m). This is further corroborated by the exchange estimates which show a clear sensitivity to the spine density. A similar pattern is observed with FWF, where the mean diffusivity is largely independent of the exchange sensitivity ($\Gamma$) while the kurtosis shows a decline. The estimated exchange rates increase with the spine density. CTI shows a higher mean diffusivity and lower kurtosis for the orthogonal DDE acquisition. The SDE-DDE contrast is sensitive to spine density, as is the microscopic kurtosis which is based on it. Overall, Fig. 5 shows that increasing dendritic spine density is associated with higher exchange rates as detected by various dMRI methods, and that these effects are observable with realistic acquisition protocols.

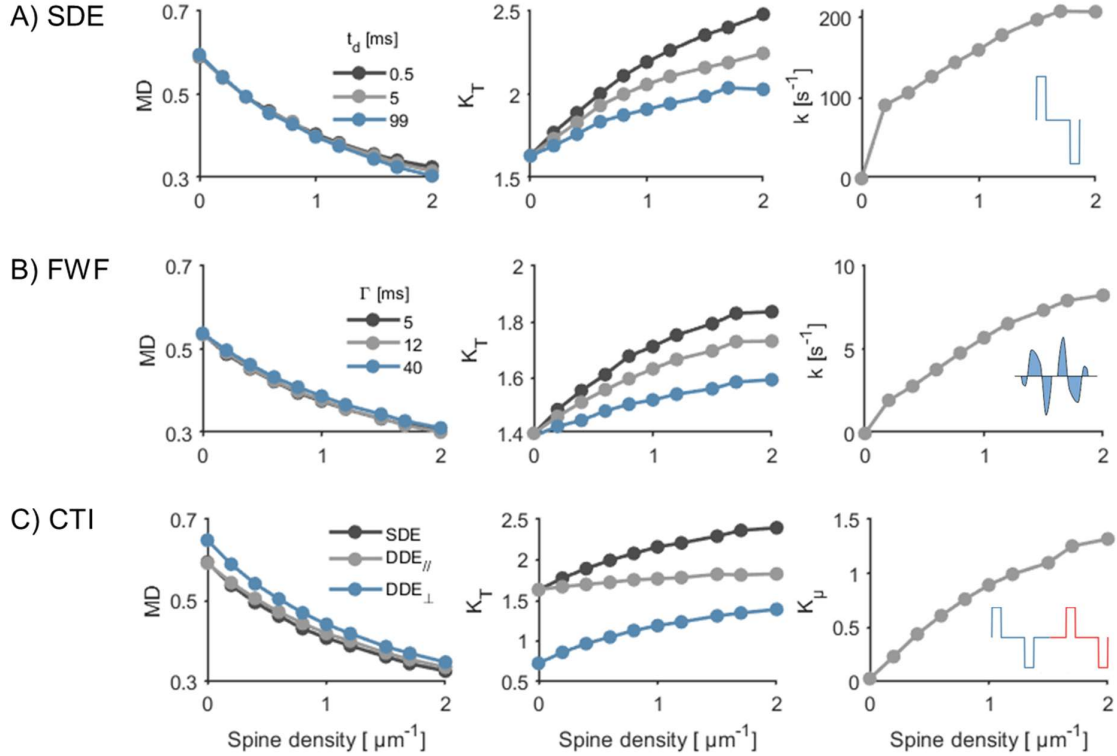

Figure 5: Impact of dendritic spines on results from three analysis methods: SDE time-dependent kurtosis, ResEx with free waveforms, and CTI. All protocols were designed for a 300 mT/m system and signals were generated considering only intracellular particles. For SDE (A), MD decreases with spine density and is predominantly independent of the diffusion time, MK increases with spine density and decreases with the diffusion time and the exchange rate grows with spine density. For FWF (B), MD is time-independent and decreases with spine density, MK increases with spine density and decays with exchange weighting. Exchange estimates grow with the spine density. For CTI (C), MD is higher for orthogonal DDE than for SDE and parallel DDE, MK follows the opposite trend, and the microscopic kurtosis grows with the spine density. The key result is that increasing spine density induces a notable impact on the exchange rate (or microscopic kurtosis) for all three dMRI methods.

Figure 6 shows mean diffusivity, kurtosis and exchange estimates obtained with SDE $K(t_d)$, ResEx (FWF) and CTI, as a function of spine density, for permeable and impermeable spines. For all three methods, permeative exchange causes an increase in MD and a decrease in kurtosis, and the dependence of the two metrics on spine density is mostly preserved. SDE exchange estimates show sensitivity to spine density in the absence of permeative exchange, but the sensitivity diminishes upon introduction of this type of exchange. FWF exchange estimates as well as CTI microscopic kurtosis increase with both permeative exchange and spine density. Overall, Fig. 6 shows that both permeative and geometric exchange influence diffusion MRI parameter estimates.

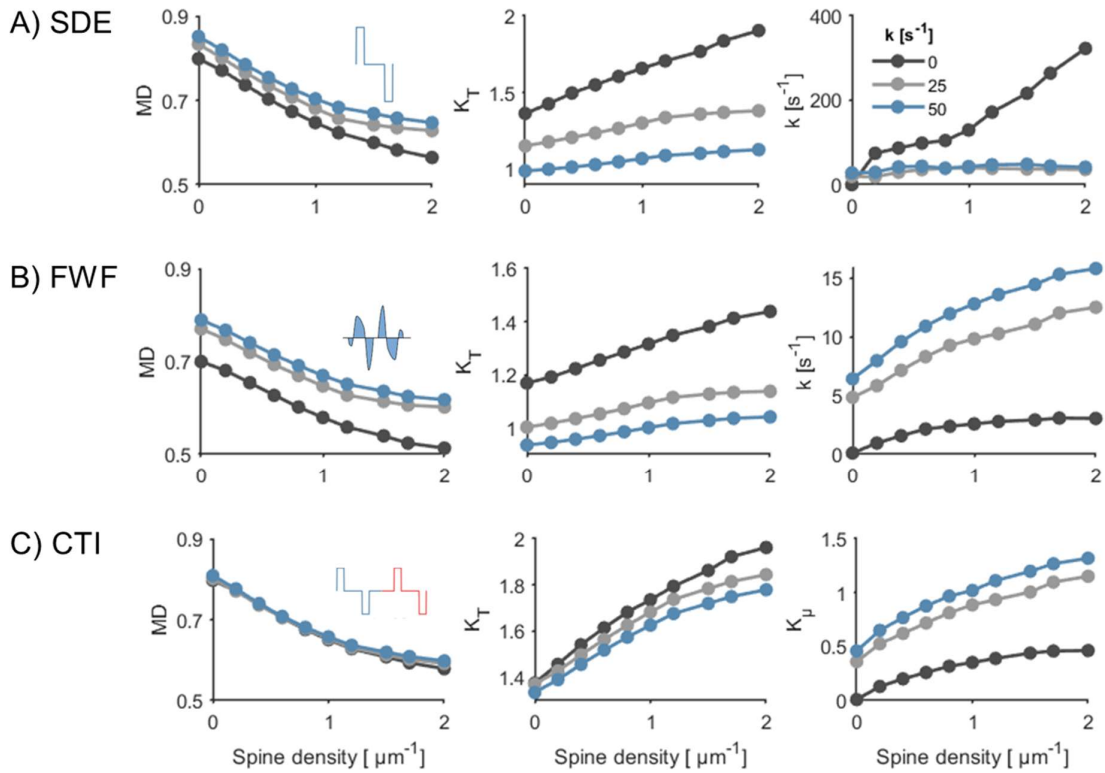

Figure 6. Influence of permeative and non-permeative exchange on diffusion MRI kurtosis and exchange estimates. Results are shown for protocols designed for a 300 mT/m system. The $k = 0$ case denotes diffusion in dendrites mixed with 20% extracellular water, while $k = 25,\ 50\ \text{s}^{-1}$ denotes the same scenario but with exchange between the two compartments at these rates. For SDE (A), MD and kurtosis are displayed for the longest diffusion time simulated. Permeative exchange increases MD and decreases kurtosis and overshadows the dependence of SDE exchange rates on spine density. For FWF (B), MD and kurtosis are shown for the longest $\Gamma$ simulated. Addition of permeative exchange elevates the MD, decreases kurtosis and increases the measured exchange rates, at all spine densities. For CTI (C), MD and total kurtosis are shown for the SDE acquisition. MD is largely unaffected by exchange, kurtosis decreases with $k$ and the microscopic kurtosis is sensitive to both permeative exchange and dendritic spines. The key message of this figure is that diffusion MRI-estimated exchange rates are influenced by both permeative and non-permeative exchange.

Kurtosis and exchange estimates obtained with tMGE are shown in Fig. 7 as a function of dendritic spine density for two values of permeative exchange rates (25 and 50 s⁻¹). MD and total kurtosis follow the same trends with spine density as observed with CTI and ResEx, except that changes in permeative exchange no longer influence these metrics. More interestingly, the transient kurtosis increases with spine density and is independent of permeative exchange, while the exchange estimates increase with permeative exchange but are independent of spine density. These trends highlight the ability of tMGE to disentangle the effects of spines (transient kurtosis) from permeative exchange. Similar trends were observed in the presence of Rician noise (Fig. A5) and for realistic diffusion encoding (Fig. A6), albeit with some fit instability at the higher exchange rate and at high spine densities. Note that while both the MD and total kurtosis in Fig. 7 also show sensitivity to spines and insensitivity to permeative exchange, we anticipate that the transient kurtosis is more specific to spines.

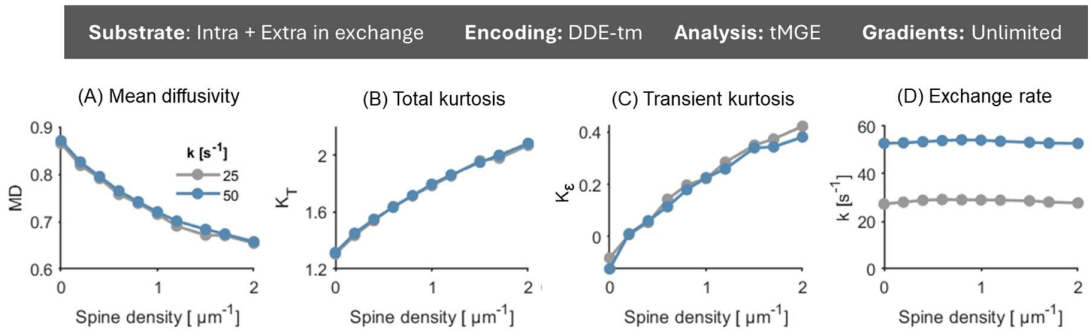

Figure 7: Separation of permeative from non-permeative exchange using diffusion MRI. Signals were generated using a multi-mixing time DDE protocol designed for arbitrary hardware constraints (Protocol VI) and analysed using tMGE. The data corresponds to diffusion in dendrites mixed with 20% extracellular water, exchanging at rates of $k = 25$ and $k = 50$ s⁻¹. Panels A and B show that MD and kurtosis depend on the spine density but are independent of permeative exchange. Transient kurtosis increases with spine density but is independent of permeative exchange. Exchange estimates are independent of spine density but increase with permeative exchange. Overall, the figure illustrates that tMGE has the potential to tease apart dendritic spines and permeative exchange.

# 5 Discussion

In this study, we investigated the influence of dendritic spines on water exchange rates in diffusion MRI using simulations. Our results highlighted that diffusion within dendritic spines imprints an exchange signature on diffusion-weighted signals, such as a signal decrease with increasing diffusion time (Fig. A1) and a decay in the kurtosis-versus-time curves (Fig. 3). Exchange rates were high despite the absence of transmembrane water transport and were sensitive to both spine density and morphology. Although the actual exchange rates we estimated were modulated by the acquisition protocol (range of diffusion times and pulse width, Fig. 4), the influence of spines could be observed regardless of the diffusion MRI analysis method used (Fig. 5). Microscopic kurtosis estimated with CTI also exhibited sensitivity to spine density—expected since it is estimated using the same signal contrast as the exchange rate. Furthermore, we explored the interplay between diffusive exchange within dendrites, and permeative exchange between dendrites and the extracellular space, and found that both processes bear a similar signature on diffusion MRI (Fig. 6). Importantly, sensitivity of the exchange rate to spine density persisted regardless of the permeability of the dendrite, suggesting that both spines and membrane permeation may impact exchange estimates in biological tissue. Finally, we made a first attempt at separating spines and permeative exchange using the recently proposed tMGE analysis framework and the results show promise (Fig. 7).

In this work, we found that dendritic spines give rise to geometric exchange inside dendrites (Fig. 2), with a geometry dependence in line with previous studies investigating diffusive leakage rates in similar structures [41,44–48]. The observed small misalignment between theory and simulated rates can be attributed to the progressive decline in particle concentration at the zone of exit, leading to time-dependent exit rates. Another explanation, relevant for the shaft-to-spine estimates, is violation of the narrow-escape condition. For realistic variations in spine morphology, geometric exchange rates between 0 and 3200 s$^{-1}$ can be expected (assuming that the total exchange rate is given by the sum of forward and reverse rates). Simulations with short pulses showed that the mapping of the geometric exchange in spines to the dMRI estimates is, however, not trivial, and is likely degenerate since different combinations of spine morphologies can yield the same exchange rate (Fig. 3). In all realisations of dendrite considered in this work, the mean diffusivity decreased rapidly with the diffusion time towards a plateau, indicating restricted diffusion and in line with previous work [31]. The kurtosis generally decreased with diffusion time—a well-known signature of exchange [14,51,53]. Exchange estimates followed the dependence on spine morphology predicted by theory, but the absolute values were generally lower (Fig. 3C). This difference could be because the MRI-measured exchange rate is not a simple sum of the spine-to-shaft and shaft-to-spine rates, or stem from higher-order terms in the cumulant expansion, or the non-mono-exponential nature of the exchange process leading to sensitivity on the timing of the acquisition protocol.

Regarding the last two points, we found that the effect of dendritic spines on dMRI estimates of water exchange rates is dependent on the acquisition protocol, specifically diffusion time range and pulse width (Fig. 4). Short diffusion times are required to probe the rapid decay of the kurtosis-vs-diffusion time curve which occurs at short times for high spine densities [34]. Protocols with long diffusion times miss this dynamic and probe the equilibrated regime (c.f supplementary Fig. A3). As has been noted by Jelescu et al. [15], variation in protocol (range of diffusion times used) is an explanation for the discrepancy between literature values of exchange rates in grey matter. The diffusion time is essentially a filter for the exchange times that can be reliably estimated. Studies using different ranges of diffusion times probe different exchange regimes, and this is an especially important consideration when the exchange estimates stem from multiple processes with potentially different characteristic timescales. Our results also indicate that the pulse width influences diffusion MRI exchange estimates in dendritic spines (Fig. 4B). Finite pulse widths manifest as a low-pass filter for the diffusion spectrum, leading to a reduction of the mean diffusivity [13]. Previous work on diffusional kurtosis estimation in exchanging systems suggests that the use of finite pulse durations influences kurtosis estimates by at most a few percent [65] and this also agrees with our results (c.f supplementary Fig. A4). However, as the current work shows, such small variations translate to a non-negligible effect on the estimated exchange rates. It should also be noted that the Kärger model, on which Eq. 13 is based, is valid in the narrow pulse regime. This in practice means that reliable exchange estimation requires the pulse width to be much shorter than the exchange time [15], which explains why the discrepancy in exchange estimates obtained with different pulse widths (Fig. 4B) is larger at higher spine densities. A final remark regarding protocol dependence is that the ultimate limiting factor for the use of narrow pulses and short diffusion times is the hardware. Reliable exchange estimation in complex media such as grey matter requires access to high-performance MRI scanners [3,4].

In this study, we observed that the impact of dendritic spines on diffusion MRI exchange estimates is evident regardless of the analysis method used (Fig. 5A-B). Estimates obtained with SDE and time-dependent kurtosis using a protocol executable on a Connectom or MAGNUS scanner were up to 200 $s^{-1}$ for spine densities up to 2 $\mu m^{-1}$. These results are within the range reported by previous work applying a similar protocol and analysis method in the healthy human brain [4]. ResEx exchange estimates were an order of magnitude lower (8 $s^{-1}$ at a spine density of 2 $\mu m^{-1}$), likely because this approach does not account for residual kurtosis resulting from powder-averaging or non-exchanging compartments in the imaging voxel. However, the estimates were in good agreement with those reported by previous work applying the same protocol and analysis method in the healthy human brain [3]. These findings provide support for the potential influence of dendritic spines on the exchange rates estimated in grey matter by these studies. Our results (Fig. 5C) also showed that the microscopic kurtosis from CTI responds to changes in dendritic spine density in the same way as the exchange rate from the other analysis methods. This is an expected result, since microscopic kurtosis is based on the same contrast as the exchange rate—the signal difference between SDE and parallel DDE at the same b-value [56,57].

Another important finding of this work is that the influence of dendritic spines on diffusion MRI-estimated exchange rates (and microscopic kurtosis) persists in the presence of permeative exchange (Fig. 6). Both the exchange rate and microscopic kurtosis respond to changes in both dendritic spine density and membrane permeability, which presents the challenge of determining which mechanism is dominant. This challenge has been highlighted in previous studies noting that structural disorder along neurites and the permeative exchange between neurites and the extracellular space are competing mechanisms behind the kurtosis time dependence observed in grey matter [15,66–68]. This is especially true when the correlation time for diffusion across the structural irregularities along neurites is of the same order as the permeative exchange time. Note that both dendritic spines and beading are classified in these studies as disorder.

Exchange effects have become increasingly evident in diffusion MRI measurements, especially in grey matter [3,8,15,16,25,29,68–71]. While most studies interpret *in vivo* exchange estimates in terms of membrane permeability, with some even converting the exchange rates to permeability via the surface-to-volume ratio [15,16], we demonstrate in this work that dendritic spines constitute an additional mechanism of exchange (Fig. 3). The effect of spines on intracellular diffusion-weighted signals has been studied before in the context of magnetic resonance spectroscopy of intracellular metabolites [31]. That simulation study focused on the mean diffusivity and its dependence on diffusion time and frequency at different spine densities. The results agree with those presented herein. Non-permeative exchange, albeit not necessarily intracellular, has also been studied recently using filter-exchange imaging in complex geometries [30]. The strength of the current work, however, is that it considers a specific non-permeative exchange mechanism (dendritic spines) and characterises—with inspiration from physiology—how alterations in spine morphology and density manifest in diffusion MRI estimates of water exchange, for a range of acquisition protocols and analysis methods.

One approach to the problem of competing exchange mechanisms is, as previously suggested by us [3], to interpret exchange as any process giving rise to a "temporal loss of observed diffusional heterogeneity", which encompasses both permeative and non-permeative exchange. Better yet, the present study teases apart the two mechanisms using the recently-proposed tMGE analysis framework [57]. The approach assumes distinct correlation times for permeative and diffusive exchange and leverages multiple-mixing-time DDE data to estimate a diffusive exchange-dependent transient kurtosis and a permeative exchange-dependent exchange rate (Fig. 7). We demonstrated the utility of the method using noiseless signals generated with arbitrary gradient strengths, but results are promising in the presence of noise at SNR = 200 as well as realistic gradient waveforms designed for a 300 mT/m-scanner (Fig A5, A6).

While both MD and MK respond to the change in spine density, we invest our efforts into estimation of transient kurtosis unconfounded by permeative exchange because this kurtosis source is more specific to spines than the other parameters. A decrease in spine density is known to be associated with neurological conditions such as schizophrenia, especially in the frontal cortex [72]. Therefore, we hypothesise that exchange rates measured with diffusion MRI could serve as a proxy for dendritic spine density, potentially enabling the early detection and characterisation of neurological diseases. Reliable mapping of dendritic spine density *in vivo* is very difficult, and the current state-of-the-art method (positron emission tomography) has low specificity and demands exposure to ionising radiation, limiting wide-spread use [73]. We predict that the application of diffusion MRI to noninvasively map dendritic spine density, if successful, would have far-reaching implications for neuroscience.

Some limitations of this work are worth discussing. While the dendritic spine model we employed in this work was sufficient to demonstrate the effects presented, it is based on regular shapes and does not reflect the realism of dendritic spines in tissue. A model more reflective of tissue would be based on 3D microscopy reconstructions of real neurons (available at neuromorpho.org) and perhaps also include somas. Furthermore, the geometry of spiny dendrites makes close-packing difficult, which necessitated an artificial simulation design to mimic exchange between dendrites and the extracellular space—again not a true reflection of biological tissue. Moreover, the methods we used to estimate exchange rate in this work are fundamentally based on the Kärger model, which is valid for barrier-limited exchange between fully coarse-grained compartments [6,14]. However, we observed time-dependence of the mean diffusivity in some geometries (Fig. 3) and exchange was likely not barrier-limited at the higher spine densities. Finally, since all analysis methods considered in this work are based on the cumulant expansion, the effects of higher order terms cannot be ruled out. Future work will perform rigorous protocol optimisation to address these shortcomings.

In conclusion, this work has shown that non-permeative exchange in dendritic spines induces time-dependence signatures on the diffusion-weighted signal mimicking permeative exchange. Dendritic spines play a non-negligible role in exchange estimates obtained with diffusion MRI. Future studies estimating exchange in grey matter should consider this mode of exchange in the interpretation of their results.

# 6 Supplementary

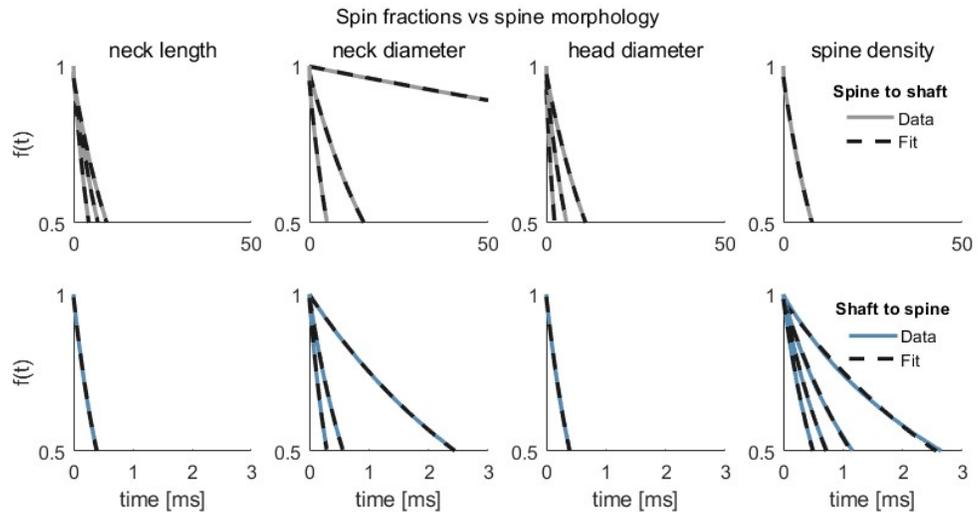

Figure A1: Temporal dynamics of spine and shaft fractions. f(t) denotes the fraction of spins remaining in the compartment (spine or shaft) after time t. These data were used to obtain the one-directional spine-to-shaft and shaft-to-spine exchange rates.

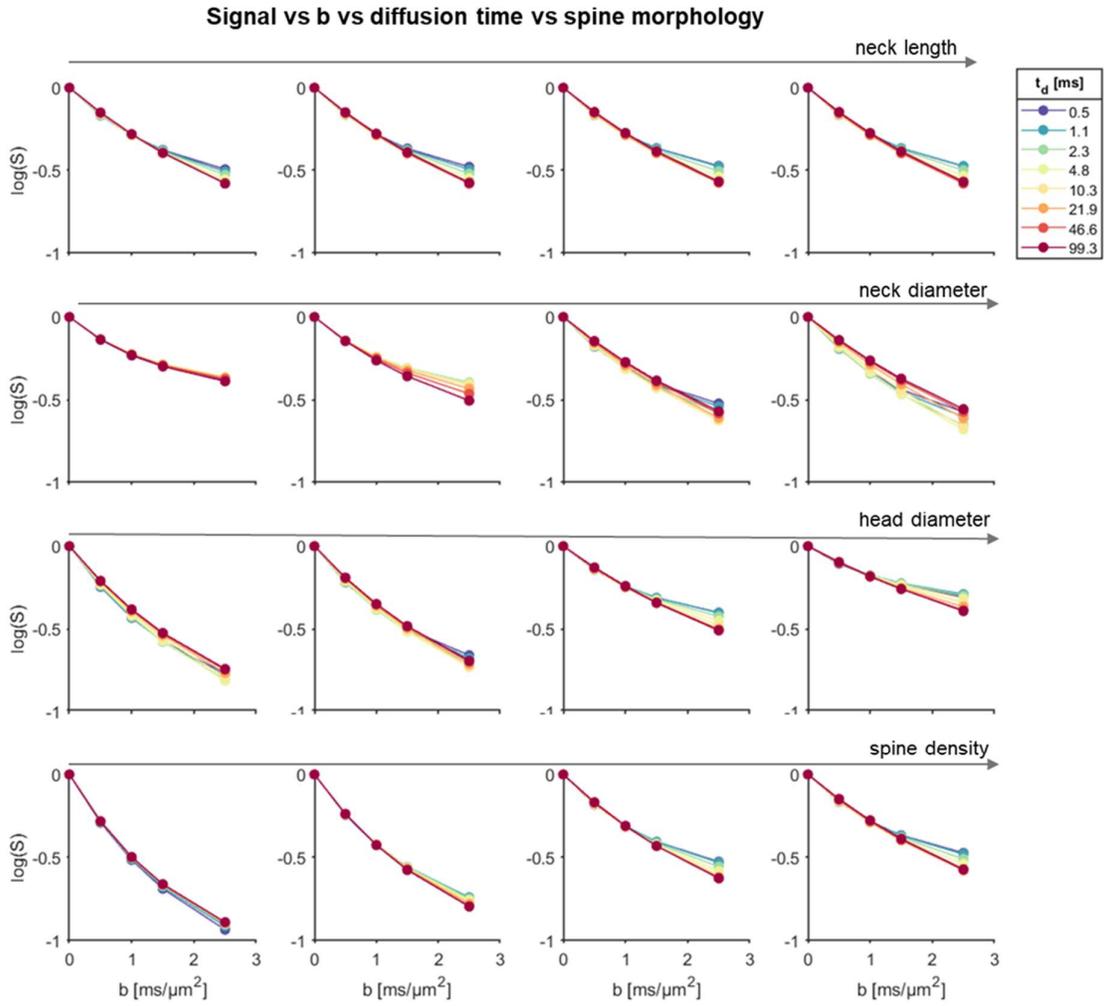

Figure A2: Diffusion time-dependence vs dendrite morphology. Each subplot shows signal vs b-value curves at different diffusion times. These curves are shown for varying neck length, neck diameter, head diameter and spine density. Signal decrease with diffusion time indicates exchange, while signal increase with diffusion time denotes restricted diffusion.

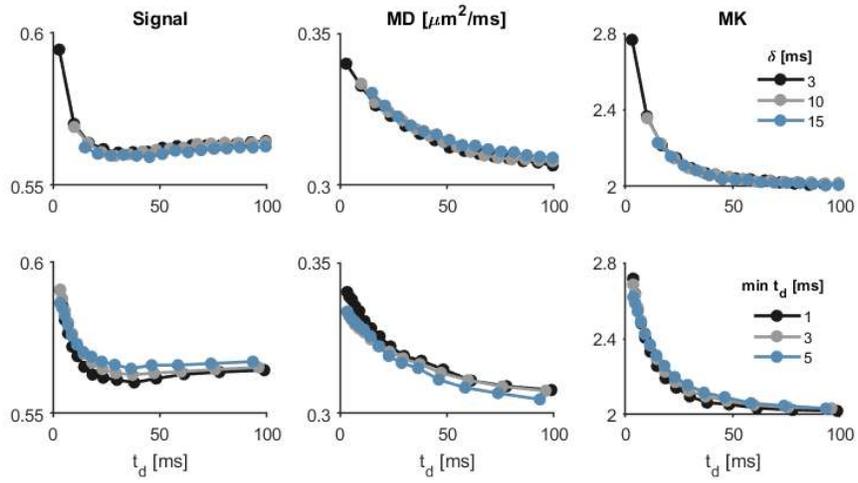

Figure A3: Signal, MD and kurtosis as a function of diffusion time and pulse width in an SDE experiment.

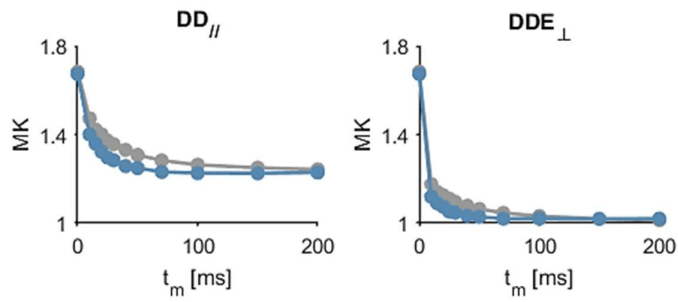

Figure A4: Mean kurtosis for parallel and orthogonal DDE as a function of diffusion time, for two exchange rates: 25 /s (grey) and 50 /s (blue). These data were used to invert the tMGE representation, enabling joint estimation of transient kurtosis and exchange rate.

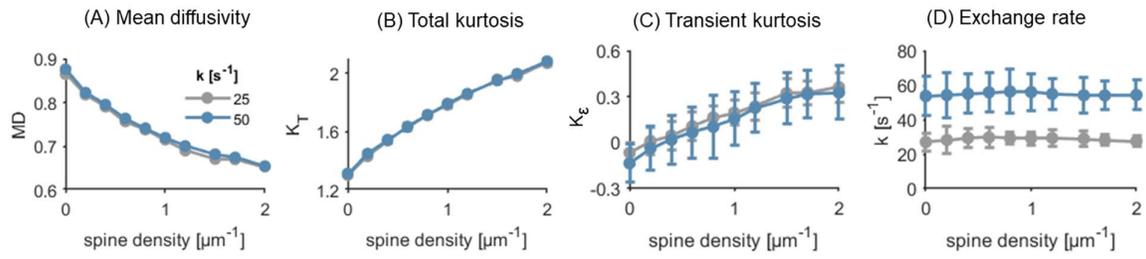

Figure A5: Equivalent of Fig. 7 in the main text, generated after corrupting the signals with Rician noise at SNR = 200.

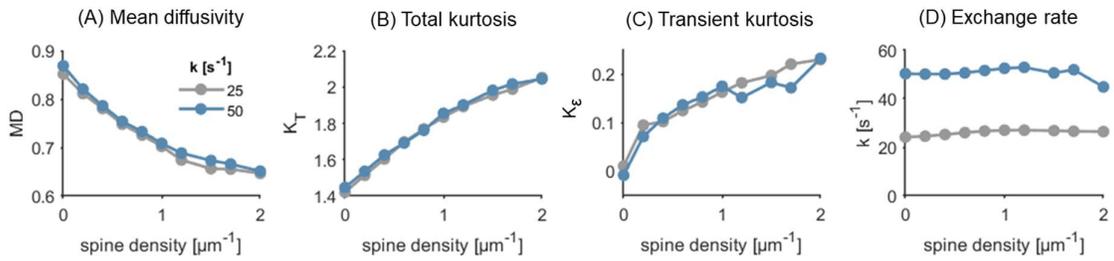

Figure A6: Equivalent of Fig. 7 in the main text, generated using gradients designed for a 300 mT/m system (Protocol VII in Table 2).

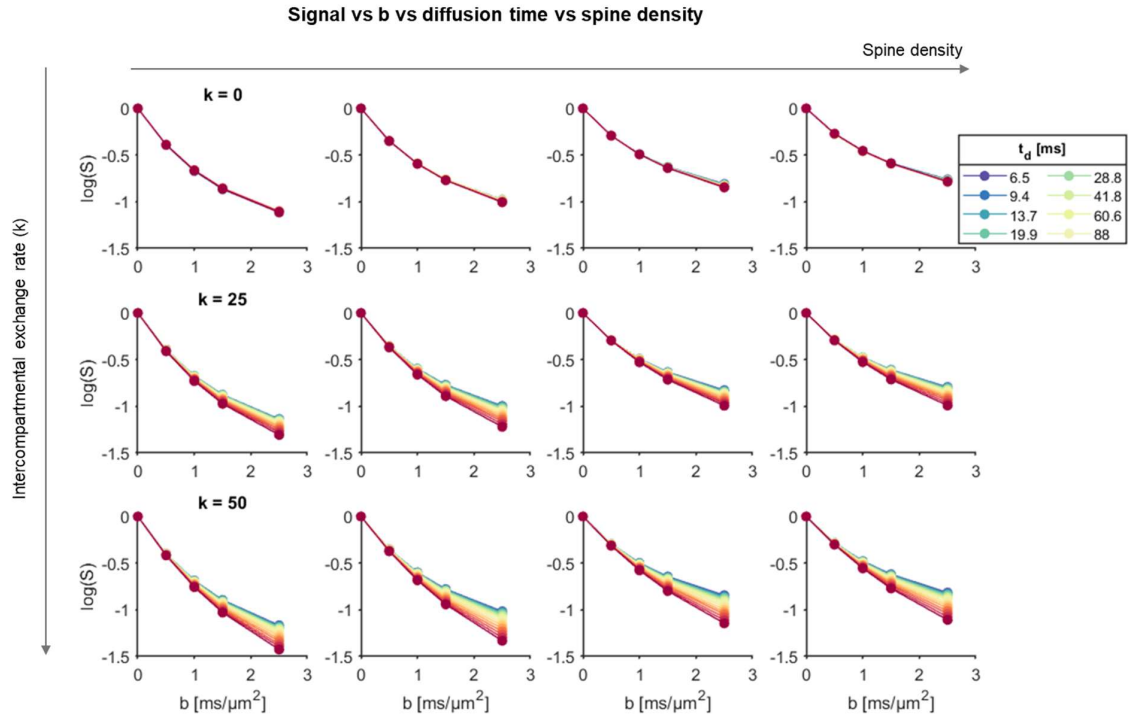

Figure. A7: SDE signal-vs-b-vs-diffusion time curves at 300 mT/m. The results are shown for varying spine density and intercompartmental exchange rate.